\documentclass[sigconf,authorversion,nonacm]{acmart}
\usepackage{fancyhdr}
\AtBeginDocument{%
    \addtolength{\footskip}{3.0\baselineskip} 

    \fancyfoot[L]{%
        \small\textit{%
        © {ACM} 2025. This is the author's version of the work. 
        It is posted here for your personal use. Not for redistribution. 
        The definitive Version of Record was published in Workshops of the International Conference for High Performance Computing, Networking, Storage and Analysis, 
        \url{https://doi.org/10.1145/3731599.3767528}%
        }%
    }

}

\usepackage{algorithmic}
\usepackage{graphicx}
\usepackage{textcomp}
\usepackage{xcolor}
\usepackage{makecell}
\usepackage[absolute,overlay]{textpos}

\begin{document}

\title{Accelerating Gravitational $N$-Body Simulations Using the RISC-V-Based Tenstorrent Wormhole\textsuperscript{\texttrademark}}

\author{Jenny Lynn Almerol}
\affiliation{%
  \institution{E4 Computer Engineering}
  \city{Scandiano}
  \country{Italy}}
\email{jenny.lynn@e4company.com}

\author{Elisabetta Boella}
\affiliation{%
  \institution{E4 Computer Engineering}
  \city{Scandiano}
  \country{Italy}}
\email{elisabetta.boella@e4company.com}

\author{Mario Spera}
\affiliation{%
  \institution{Scuola Internazionale Superiore di Studi Avanzati}
  \city{Trieste}
  \country{Italy}}
\email{mspera@sissa.it}

\author{Daniele Gregori}
\affiliation{%
  \institution{E4 Computer Engineering}
  \city{Scandiano}
  \country{Italy}}
\email{daniele.gregori@e4company.com}

\begin{abstract}
Although originally developed primarily for artificial intelligence workloads, RISC-V-based accelerators are also emerging as attractive platforms for high-performance scientific computing. In this work, we present our approach to accelerating an astrophysical $N$-body code on the RISC-V-based Wormhole n300 card developed by Tenstorrent. Our results show that this platform can be highly competitive for astrophysical simulations employing this class of algorithms, delivering more than a $2 \times$ speedup and approximately $2 \times$ energy savings compared to a highly optimized CPU implementation of the same code.
\end{abstract}



\maketitle

\begin{textblock*}{\paperwidth}(18mm,0.95\paperheight)
    \parbox{\paperwidth}{
        \small\itshape
        © ACM 2025. This is the author's version of the work. 
        It is posted here for your personal use. Not for redistribution. The definitive Version of Record was published in

        Workshops of the International Conference for High Performance Computing, Networking, Storage and Analysis, 
        \url{https://doi.org/10.1145/3731599.3767528}
    }
\end{textblock*}

\section{Introduction}
The royalty-free, open-standard Instruction Set Architecture RISC-V is gaining widespread popularity due to its key features: open-source nature, community-driven development, modularity, and extensibility~\cite{patterson_book}. While early RISC-V platforms primarily targeted embedded systems, recent advancements have attracted significant interest from the High Performance Computing (HPC) community, driven by the development of 64-bit processors, vector extensions, and specialized accelerators fueled by the growth of Artificial Intelligence (AI)~\cite{venieri2025montecimonev2road,Teresa,MonteCimone}. Although RISC-V-based accelerators were initially designed with AI and machine learning (ML) workloads in mind, their hardware capabilities are well-suited to accelerate linear algebra operations, which underpin many scientific applications~\cite{brown2025exploring,brown2024accelerating}. One notable RISC-V-based accelerator earning traction in HPC is the Wormhole n300 card developed by Tenstorrent~\cite{TenstorrentWormhole}. By effectively decoupling computation from data movement, this accelerator offers the potential for high performance and energy efficiency at a relatively modest cost~\cite{brown2025exploring}.

Many scientific disciplines rely heavily on HPC, leveraging supercomputers and parallel computations; astrophysics is a prime example~\cite{SPACE1,SPACE2}. Numerical simulations play a crucial role in modeling and interpreting increasingly detailed observational data. Therefore, it is essential to assess whether RISC-V architectures can provide a viable solution in this domain by ensuring both performance and energy efficiency. To this end, in this study, we focus on offloading the most computationally expensive calculations of our home-grown $N$-body simulation code to the RISC-V-based Tenstorrent Wormhole accelerator. We then systematically assess its performance and energy efficiency relative to traditional CPU implementations.

Overall, the contributions of this paper are:
\begin{itemize}
\item We port our 
$N$-body simulation code to the RISC-V-based Tenstorrent Wormhole accelerator; to the best of our knowledge this is the first example of an astrophysical application leveraging a RISC-V accelerator;
\item We evaluate the performance of the accelerated code in terms of computational time and energy consumption, and compare these results against a reference CPU implementation of the same algorithm.
\end{itemize}

This paper is structured as follows. Section \ref{sec:background} provides an overview of our in-house $N$-body application, the Wormhole n300 card, and the TT-Metalium programming interface used to offload computations to the Wormhole device. Section \ref{sec:strategy} describes our approach to accelerate the code. Section \ref{sec:experiments} presents our experiments and results. Finally, section \ref{sec:conclusion} concludes the paper.

\section{Background}\label{sec:background}
Direct gravitational $N$-body simulations numerically solve the equations of motion for systems of $N$ particles by explicitly computing the mutual gravitational forces between all pairs. Because they do not rely on approximations, these simulations are essential for accurately modeling dense stellar systems, such as star clusters \cite{spurzem1999direct}. These systems are considered to be the primary environments for the formation of compact object binaries, such as black hole binaries. The coalescence of these binaries produces gravitational waves detectable by current observatories, including LIGO, Virgo, and KAGRA \cite{abbott2020prospects}, as well as by future detectors like the Einstein Telescope \cite{einstein_telescope}.  With upcoming instruments expected to detect millions of sources, it has become increasingly important for the astrophysical interpretation of gravitational waves to develop \emph{efficient} and \emph{scalable} direct $N$-body codes.

In this study, we port and optimize our $N$-body code to execute efficiently on a RISC-V–based accelerator. The open and modular nature of the RISC-V architecture enables customization of compute platforms, which is crucial for meeting the energy efficiency and scalability demands of large-scale scientific simulations like $N$-body problems. Among the emerging RISC-V-based accelerators, the Tenstorrent Wormhole stands out as a notable example. Designed for AI/ML workloads, it operates at up to $160 \, \mathrm{W}$ and offers a scalable, cost-effective alternative to traditional Graphics Processing Units (GPUs) \cite{TenstorrentWormhole}. Its architecture is optimized for scalability and efficient data movement, both of which are essential for managing the immense data and communication demands of large-scale $N$-body simulations.

\paragraph{Tenstorrent Wormhole}

The Wormhole is an Application-Specific Integrated Circuit (ASIC) organized as a grid of processing tiles. Each chip integrates 64 programmable processing elements known as Tensix cores, along with additional RISC-V control processors \cite{tenstorrent_isa_docs}. It connects to $12 \, \mathrm{GB}$ of external GDDR6 memory via a 192-bit memory bus, providing high-bandwidth access to off-chip data \cite{TenstorrentWormhole}.
For high-throughput communication, the design includes two QSFP-DD ports capable of bidirectional data transfer at up to $200 \, \mathrm{Gbps}$, 
and a PCIe 4.0 x16 interface for host connectivity. Each Ethernet core (ERISC) integrates a RISC-V processor, $256 \, \mathrm{kB}$ local cache, and an Ethernet subsystem for high-speed data transmission \cite{TenstorrentWormhole, corsix_tenstorrent_series}.

\begin{figure}[htb!]
    \centering
    \includegraphics[width=1.0\linewidth]{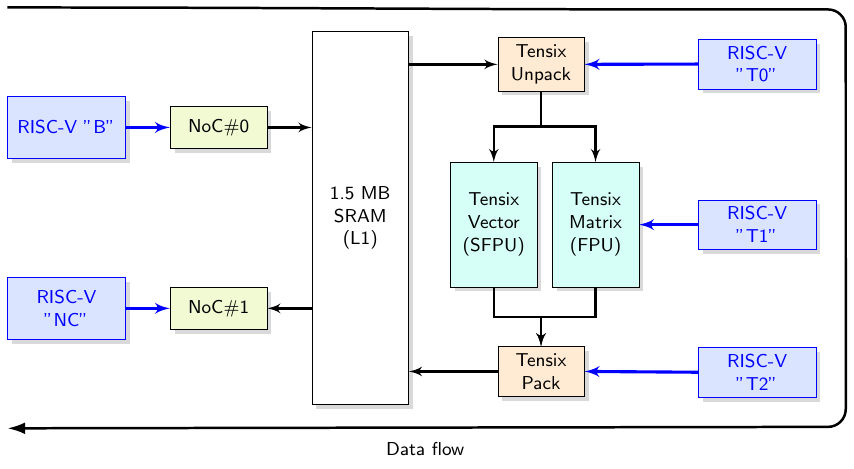}
    \vspace{-4mm}
    \caption{Simplified schematic overview of a Tensix core in the Tenstorrent Wormhole AI accelerator. Blue arrows correspond to instruction dispatch, while black arrows indicate data movement \cite{chang2025tensix}.}
    \label{fig:tensix_architecture}
\end{figure}

Tensix cores, illustrated in Fig.~\ref{fig:tensix_architecture}, provide the system main compute horsepower designed for efficient matrix and vector computations. Each core comprises (a) five embedded ``Baby'' RISC-V CPUs for control and coordination, (b) a high-throughput tensor math unit (Floating Point Unit, FPU) for low-precision matrix arithmetic, (c) a wide SIMD engine for general-purpose vector operations (Scalar Floating Point Unit, SFPU), 
(d) $1.5 \, \mathrm{MB}$ of SRAM (L1), and (e) two network routers interfacing with Networks-on-Chip (NoCs) \cite{tenstorrent_isa_docs,corsix_tenstorrent_series}.
The NoC serves as a scalable communication backbone, allowing tiles to efficiently exchange data and access memory across the chip. Through NoC transactions, any tile can initiate read or write operations on the memory located on another tile. The “Baby" RISC-V core, on the other hand, is a lightweight 32-bit, in-order, single-issue processor, optimized for area and power efficiency, and operates at a clock frequency of $1 \, \mathrm{GHz}$~\cite{corsix_tenstorrent_series}.

As shown in Fig.~\ref{fig:tensix_architecture}, the five “Baby” RISC-V cores within each Tensix processor are functionally divided into two data movement cores (RISC-V NC and B) and three compute cores (RISC-V T0, T1, and T2).
The data movement cores are responsible for coordinating data transfers between the Tensix core and off-chip DRAM, while the compute cores coordinate arithmetic and logic operations by issuing instructions to various Tensix coprocessor units, including the tensor FPU and the SFPU.

Within the Tensix compute pipeline, the three compute cores assume specialized roles. Although any Tensix execution resource can, in principle, be driven by any of the three compute cores and any core can read from or write to either NoC router, the diagram referenced here adopts the traditional mapping. RISC-V T0 (UNPACK) issues instructions to the unpacker module, which loads data from SRAM into two $4\, \mathrm{KiB}$ source registers, \verb|srcA| and \verb|srcB|. Each of these registers are capable of holding up to 1024 single-precision floating-point values. RISC-V T1 (MATH) controls the main arithmetic datapath, issuing instructions to the ThCon, SFPU, and FPU units to perform operations on the contents of \verb|srcA| and \verb|srcB|. Finally, RISC-V T2 (PACK) directs the movement of results from the $32 \, \mathrm{KiB}$ destination register, \verb|dst|, organized into 16 segments, back into SRAM~\cite{deepwiki_tensix}. In the Tenstorrent Software Development Kit (SDK) TT-Metalium, the hardware-level partitioning is reflected in the software execution model, wherein data movement cores execute data movement kernels, while the compute cores are assigned specialized roles in executing compute kernels~\cite{tenstorrent_isa_docs}.

\paragraph{Tenstorrent TT-Metalium}
TT-Metalium is a low-level programming interface that exposes direct control over Tenstorrent hardware, enabling the construction of highly parallel, fine-grained compute pipelines. The standard development workflow begins with the initialization and configuration of the hardware device, including the setup of communication channels between the host and the device. Subsequently, compute kernels are compiled and loaded onto the device, with execution parameters defined explicitly. Memory buffers are then allocated, and data is transferred between the host and device to prepare for computation. Finally, kernels are enqueued for execution via a command queue, which manages dispatch, synchronization, and sequencing of tasks on the hardware~\cite{deepwiki_tensix, tenstorrent_isa_docs}.

The application requires the implementation of three custom kernels (read, compute, and write) each responsible for a distinct stage of the pipeline. These kernels are executed across data movement and compute cores in a dataflow-driven manner, communicating via software-managed circular buffers (CBs). This model enables the overlap of computation and communication, as data is produced and consumed asynchronously across pipeline stages.
Utilizing TT-Metalium support for tilized tensors further enhances this pipeline by arranging data into $32 \times 32$ tiles that are contiguous in memory, enabling efficient, high-bandwidth data transfers over DRAM, NoC, and Ethernet~\cite{brown2025exploring,glagovich2025flashattention, tenstorrent_isa_docs}.

Synchronization between kernels is handled through TT-Met\-a\-li\-um buffer control primitives: 
\verb|cb_wait_front|, 
\verb|cb_pop_front|, 
and \linebreak
\verb|cb_reserve_back|. 
The first two manage consumer-side synchronization, ensuring a kernel waits for sufficient data and consumes it in order. \verb|cb_reserve_back| complements this on the producer side by blocking writes until enough buffer space is available, preventing overwrites and enforcing back-pressure. Once space is reserved, the producer writes data and calls \verb|cb_push_back| to finalize the insertion. Together, these primitives coordinate producer–consumer execution, preserve data dependencies, and prevent race conditions~\cite{brown2025exploring,glagovich2025flashattention,tenstorrent_isa_docs}.

\section{Strategy to port our $N$-body application to Tenstorrent}\label{sec:strategy}

Direct $N$-body simulations solve Newton’s equations of motion by computing the gravitational force on each particle due to all other particles in the system: 
\begin{displaymath}
  \mathbf{F}_{i} = \sum_{\substack{j=1 \\ j \neq i}}^{N} G \frac{m_i m_j}{r^3_{ij}}(\mathbf{r}_j - \mathbf{r}_i),
\end{displaymath}
where $m_i$ and $m_j$ are the particle masses, $\mathbf{r}_i$ and $\mathbf{r}_j$ are their position vectors, and $G$ is the gravitational constant. This requires $O(N^2)$ pairwise interactions per time step. This brute-force method, while highly accurate, is computationally intensive and is a major challenge for large-scale astrophysical systems due to the nested double \verb|for|-loop structure at the core of the force computation. This section provides details on the strategy to accelerate this computation by mapping it onto the tile-based architecture of the Tenstorrent Wormhole processor. This includes the calculation of the acceleration and its first time derivative (jerk) for each particle in all three spatial dimensions.

\begin{figure}[htb!]
    \centering
    \includegraphics[width=1.0\linewidth]{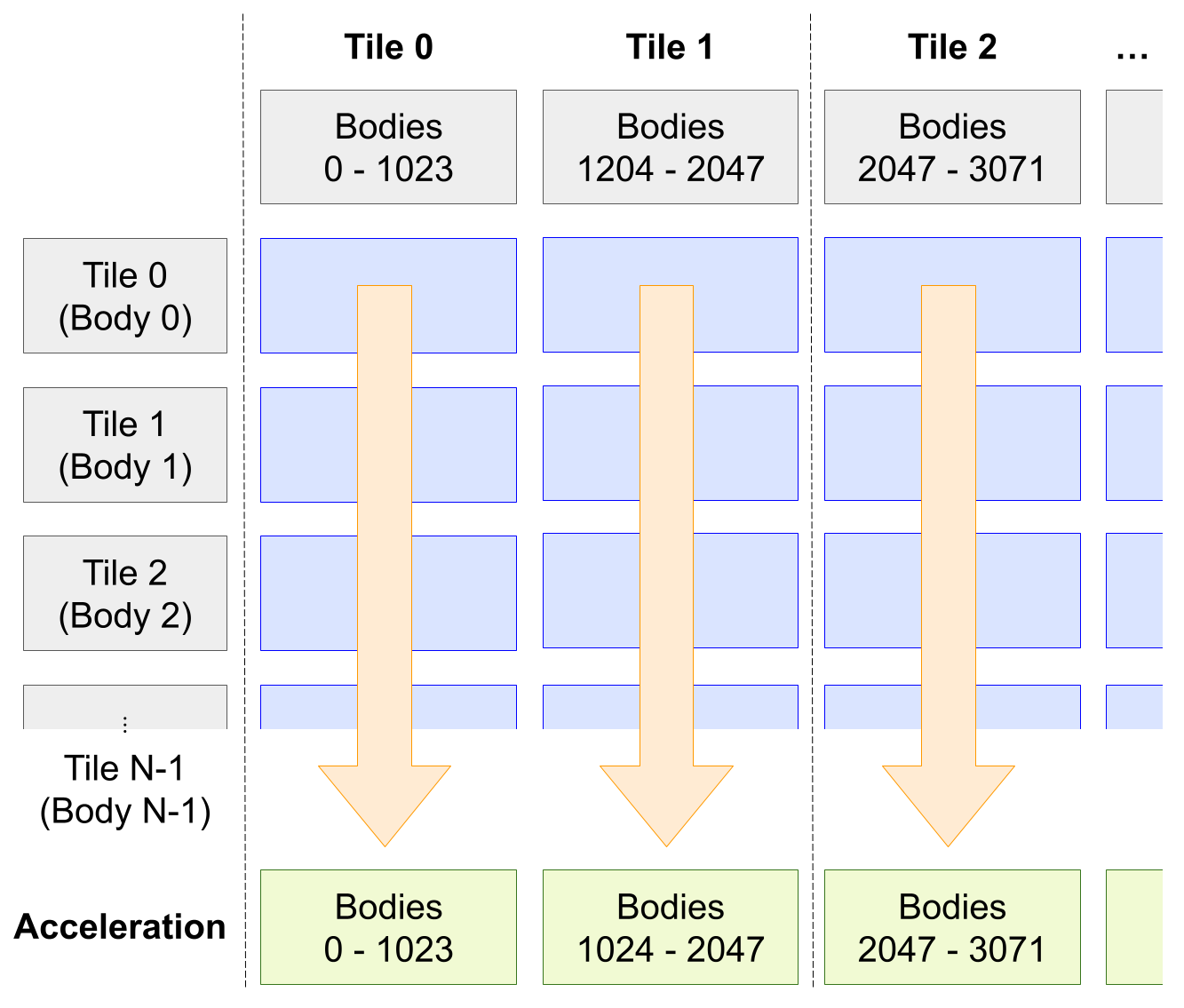}
    \vspace{-4mm}
    \caption{Tile-based parallel force calculation. Tiles within the vertical lines show the first batch of calculations when, for example, 2 Tensix cores are used.}
    \label{fig:nbody_tiles}
\end{figure}

To achieve parallelization, the outer \verb|for|-loop of the force calculation is distributed across multiple Tensix cores. Each core is assigned a subset of particles for which it computes the net gravitational force. However, to calculate this net force, each core requires access to the complete set of particle data. Therefore, we create copies of the data, organized into $N$ tiles, where each tile holds 1024 elements. This design allows the cores to consume data efficiently within the inner loop of the force calculation. Fig.~\ref{fig:nbody_tiles} illustrates this process, where the column tiles are distributed across Tensix cores, and a row represents computations done in parallel. Acceleration values are accumulated in the direction of the arrows, and once completed, are written to the output CBs.

The data flow is organized across three compute kernels. The \verb|read| kernel loads the original particle data from DRAM and formats it into tiles stored in CBs. It is implemented as a double \verb|for|-loop, where the outer loop reads the particle data in a tiled manner, and the inner loop reads the replicated tiles used in the subsequent computation. The \verb|compute| kernel then performs the gravitational force and jerk calculations by consuming the tiled data in a manner consistent with the \verb|read| kernel. After the computation is complete, the \verb|write| kernel transfers the results back to DRAM.

A Tensix core \verb|dst| register has a capacity of 16 tiles when using BFP16 data format, which is effectively halved when we utilize the FP32 format.
Given this limited register space, intermediate values that are frequently reused within the force calculation kernel are staged in on-chip SRAM using CBs. This approach is essential for managing temporary data such as the displacement vector components ($dx$,$dy$,$dz$) without causing register spills. Moreover, the arithmetic and transcendental operations inherent in the force calculation are executed on the core SFPU. We employ the TT-Metalium API to invoke hardware-accelerated functions for element-wise tile operations such as \verb|sub_binary_tile()|, \verb|square_tile()|, and \verb|rsqrt_tile()|.

This strategy leverages the architectural features of the Tenstorrent Wormhole processor to enable efficient and scalable computation of $N$-body interactions, laying the foundation for high-performance astrophysical simulations on this platform. The full implementation can be found in the Github repository\footnote{N-Body simulation source code: \url{https://github.com/jlalmerol/N-Body-Code.git}}.

Since the Tenstorrent Wormhole accelerator supports up to FP32, a mixed-precision approach is adopted where acceleration, jerk, and other intermediate values within the force calculation are computed in single precision, while all remaining calculations are performed in double precision on the CPU cores to maintain the overall numerical accuracy of the simulation. 
Such mixed-precision schemes have previously been shown to provide significant gains in performance and energy efficiency while maintaining accuracy~\cite{micikevicius2018mixed}.

To ensure the correctness of the ported $N$-Body code, force and jerk values computed by the Tenstorrent Wormhole processor are compared against a naive, double-precision brute-force implementation of the $O(N^2)$ algorithm executed on a conventional CPU. This CPU-based calculation serves as the ``golden reference" for accuracy. We ensure that discrepancies are within acceptable tolerance levels for floating-point arithmetic, with each acceleration and jerk component within $0.05 \%$ and $0.2 \%$ of a typical force magnitude, respectively, relative to the double-precision result, thereby confirming the correctness of the TT-Metalium implementation and the underlying hardware operations.

Wormhole-based $N$-body simulations are also benchmarked against an optimized, CPU-only reference implementation executed on the same host system to ensure consistency in measurements.

The reference implementation is written in C++, also in mixed precision, and parallelized using established parallel programming models such as the Message Passing Interface (MPI) and OpenMP. Additionally, it leverages AVX-512 intrinsics to efficiently compute the force between particles, maximizing single-core performance. 

The host system used to validate the accelerated code and conduct the experimental campaign is equipped with a dual-socket AMD EPYC 9124 processor, offering a total of 64 hardware threads (2 sockets $\times$ 16 cores $\times$ 2 threads per core) and a maximum clock frequency of $3.71 \, \mathrm{GHz}$.  The machine is equipped with $1.5 \, \mathrm{TB}$ of DDR5 RAM and runs the 64-bit Ubuntu 24.04.2 Linux operating system with kernel version 6.8.0.
Four Wormhole n300 devices are connected to the host via PCIe Gen 4; however, our current experiments utilize only one device.

The reference code is built via CMake using GNU Compiler Collection version 13.3.0, conforming to the C++20 standard. Distributed parallelism is facilitated through Open MPI version 4.1.6. The compilation process incorporates optimization and performance-enhancing flags, including \verb|-O3|, \verb|-fopenmp|, \verb|-march=native|, and \verb|-mavx512f|. 
The same software stack, alongside the latest TT-Metal library (v0.60.1), is used to compile the accelerated code, with flags \verb|-O3|, \verb|-fopenmp|, and \verb|-march=x86-64-v3|.

\section{Experimental Campaign}\label{sec:experiments}

We assess the performance of the Wormhole-based 
$N$-body code by measuring the duration of a representative simulation in seconds (time-to-solution) and the energy consumed by accelerators and processors to complete it (energy-to-solution).
These metrics are then compared with those of the reference implementation.
For this work, the representative simulation models 102400 particles evolving over ten time cycles. 

Before starting the simulation, we perform a device reset and surround the actual simulation with a 120-second \verb|sleep| period both before and after to allow the system to relax to idle conditions.
This workflow is typically repeated multiple times per simulation to gather sufficient statistical data.

\begin{figure}[]
    \centering
    \includegraphics[width=\linewidth]{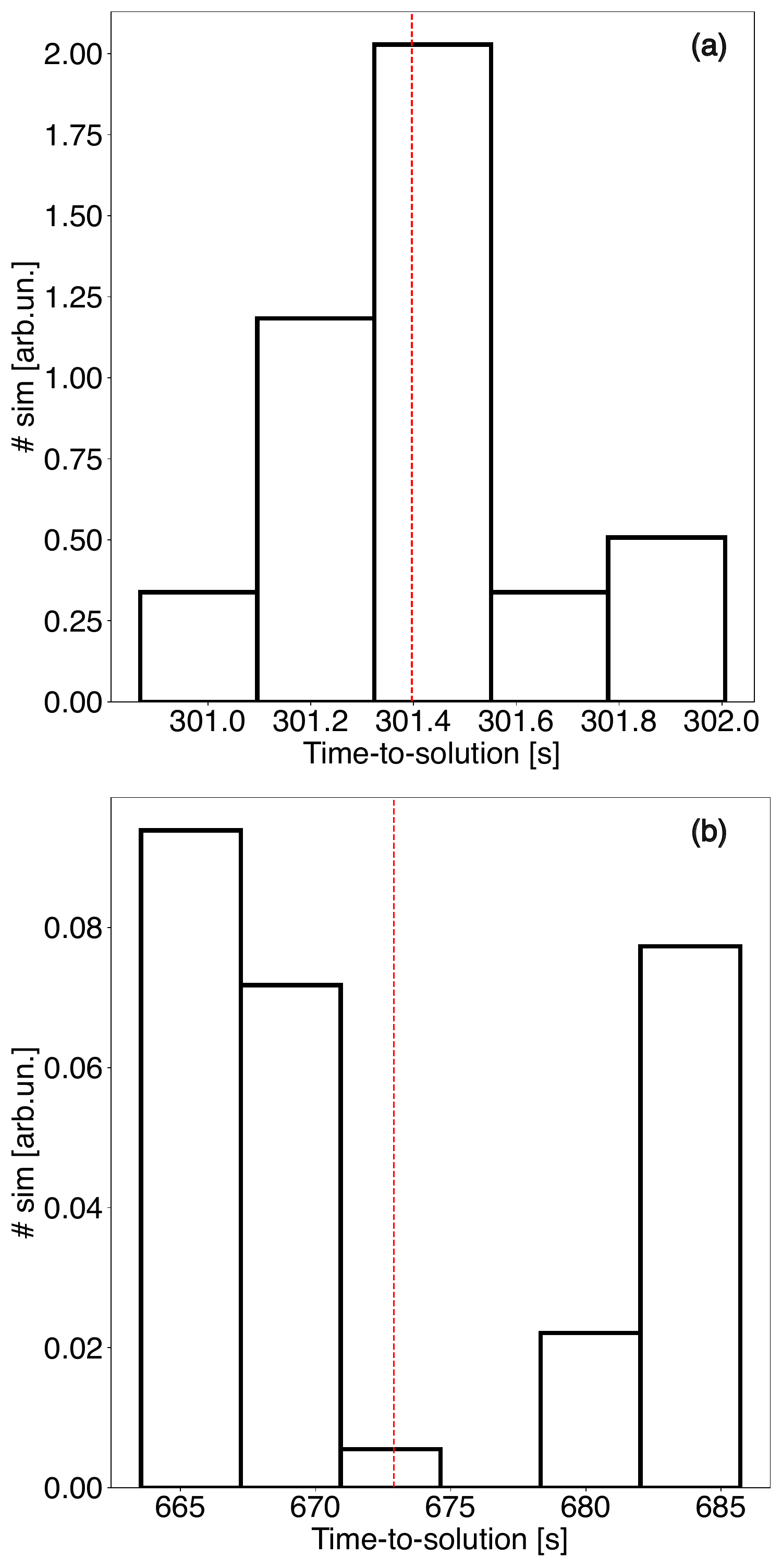}
    \vspace{-4mm}
    \caption{Distribution of time-to-solution values across simulations for runs on device + CPU (a) and on CPU only (b). The red dashed lines represent the average time-to-solution.}
    \label{fig:time_distribution}
\end{figure}

Time-to-solution is measured using hardcoded \verb|MPI_Wtime()| calls at the start and end of the simulation; this measurement does not include any time spent in \verb|sleep|.
Figs.~\ref{fig:time_distribution} (a) and (b) present histograms of the time-to-solution for simulations executed with the accelerated and reference codes, respectively.
In total, the plots report results from 26 accelerated simulations and 49 reference simulations.
Although 50 accelerated simulations were submitted using a single Wormhole card, only 26 completed successfully; the remaining 24 failed to start due to errors occurring during the device reset phase.
This issue affected all devices and is currently under investigation.
The accelerated simulations are executed with a single OpenMP thread and one MPI task, whereas the reference simulations employ 32 OpenMP threads and one MPI task.
In the latter case, all physical cores of the host are utilized by setting \verb|OMP_PLACES=cores|; using all hardware threads did not yield any significant performance improvement. 
Wormhole-based $N$-body simulations achieve shorter times-to-solution than their CPU-based counterparts.
On average, runs leveraging the Tenstorrent card completed in $301.40 \pm 0.24 \, \mathrm{s}$, compared to $672.90 \pm 7.83 \, \mathrm{s}$ for the CPU implementation, corresponding to an average speedup of $2.23 \times$.
We note that time-to-solution for CPU-based simulations exhibits a higher standard deviation.
This is likely due to variability in system load, resource contention, and operating system scheduling, which can introduce fluctuations not typically present in the more dedicated accelerator environment.

Energy-to-solution estimates are derived by combining raw and post-processed data sampled in user space at the frequency of $\approx 1 \, \mathrm{Hz}$ throughout the entire duration of a job (e.g., the simulation plus the \verb|sleep| periods)~\cite{amati}.

To calculate the energy consumed by the Wormhole cards during computation, we record the power usage of the four accelerators at roughly one-second intervals using the manufacturer system management interface \verb|tt-smi|.
The energy dissipated by the two CPUs is measured using Intel’s Running Average Power Limit (RAPL) interface, which on our AMD system exposes the energy of the two CPU Packages and of the two CPU cores (see~\cite{RAPL} for definitions).
We use two different methods to access RAPL data: we directly read RAPL registers every second and we employ \verb|perf stat -a -e| in combination with one second \verb|sleep|. We have verified that both approaches yield equivalent results, except in cases where register overflows occur. In this work, we choose the second method to avoid dealing with overflow corrections.

Additionally, in our experiments we also monitor the total server power consumption at the same frequency using \verb|ipmitool| \verb|dcmi| \verb|power| \verb|reading|.
However, we exclude this data from our analysis due to the elevated power usage of the temporary host server, which is a $4 \, \mathrm{U}$ system designed to accommodate multiple high-end GPUs and, therefore, having a high baseline power consumption.

Since data acquisition occurs in user space, we modified the \verb|kernel.perf_event_paranoid| setting to -1 and adjusted permissions for accessing RAPL and Intelligent Platform Management Interface (IPMI)-related files. 

\begin{figure}[]
    \centering
    \includegraphics[width=1.0\linewidth]
    {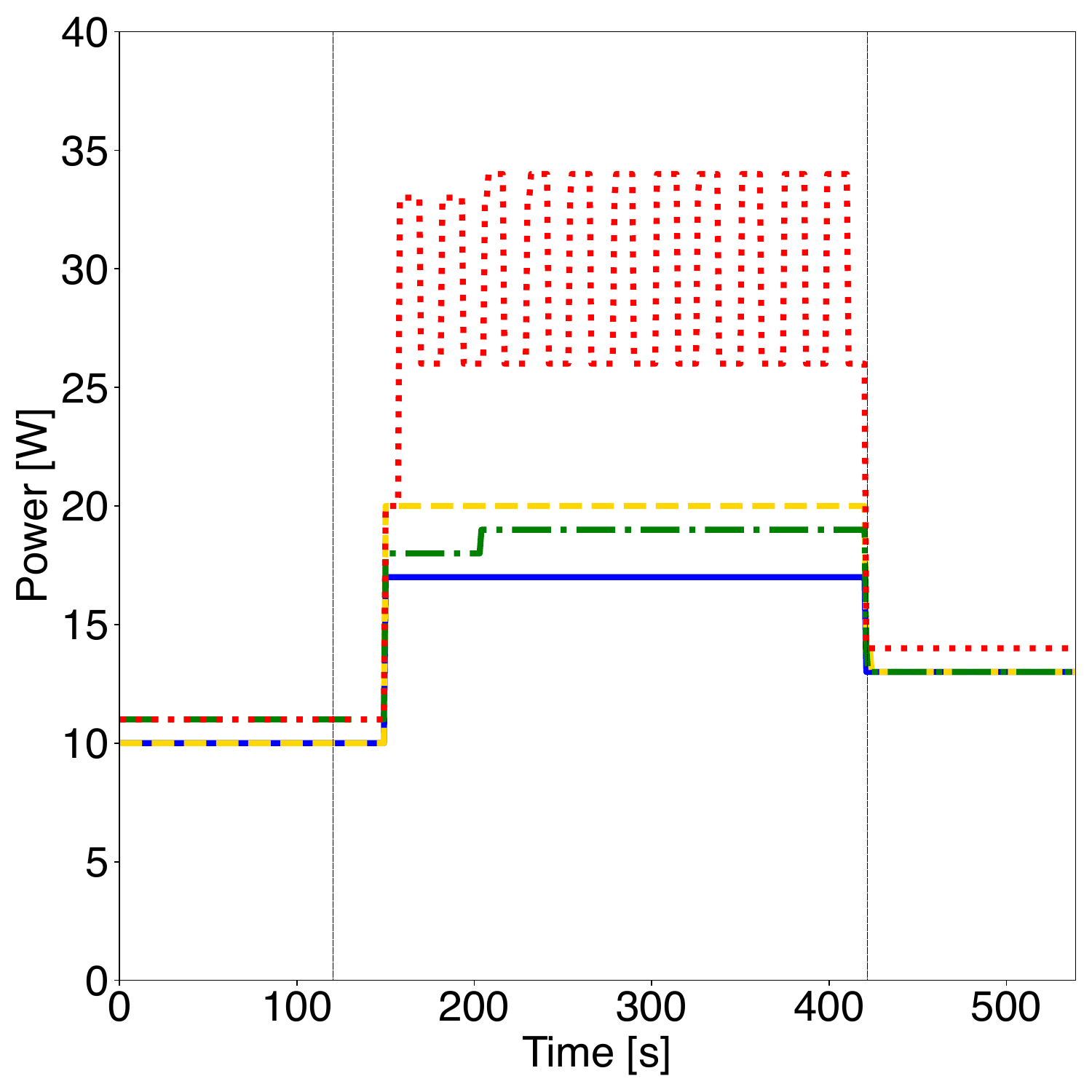}
    \vspace{-4mm}
    \caption{Time series of power absorbed by the four Tenstorrent cards during a representative run. The solid blue line corresponds to device 0, the dashed yellow line to device 1, the dash-dotted green line to device 2, and the dotted red line to device 3. Vertical solid black lines indicate the start and end of the numerical simulation.} 
    \label{fig:card_power}
\end{figure}
All sampled values are stored in csv files along with their corresponding timestamps.
Fig.~\ref{fig:card_power} shows the time series of the power absorbed by the four Tenstorrent cards during one job.
Power usage is recorded for all four cards, despite the fact that in this phase our simulations use only one device.
While idle, before the simulation starts, the cards consume between $10$ and $11 \, \mathrm{W}$.
At the beginning of the simulation, an initialization phase runs solely on the host; during this phase, the power consumption of the cards remains at idle levels.
Once the kernel responsible for computing the forces between particles is invoked, the power consumption of all four devices increases, even though only device 3 is actively used in this run. This occurs because all four devices are powered on.
The unused devices maintain a steady power consumption below $20 \, \mathrm{W}$, while the active device shows fluctuations between $26$ and $33 \, \mathrm{W}$.
These power peaks correspond to periods of intensive computation on the accelerator, whereas the lower values occur when calculations that are not offloaded are handled by the host CPU.
Once the simulation ends, the power consumption of all four cards drops sharply, returning to values similar to, but not exactly equal to, those recorded at the start of the job. This minor difference in idle power consumption, which is currently under investigation, resolves upon resetting the cards.
Although the results in Fig.~\ref{fig:card_power} were obtained using a single OpenMP thread and one specific device, the observed trends are consistent across all runs, including those using multiple OpenMP threads and different devices.

\begin{figure}[]
    \centering
    \includegraphics[width=\linewidth]{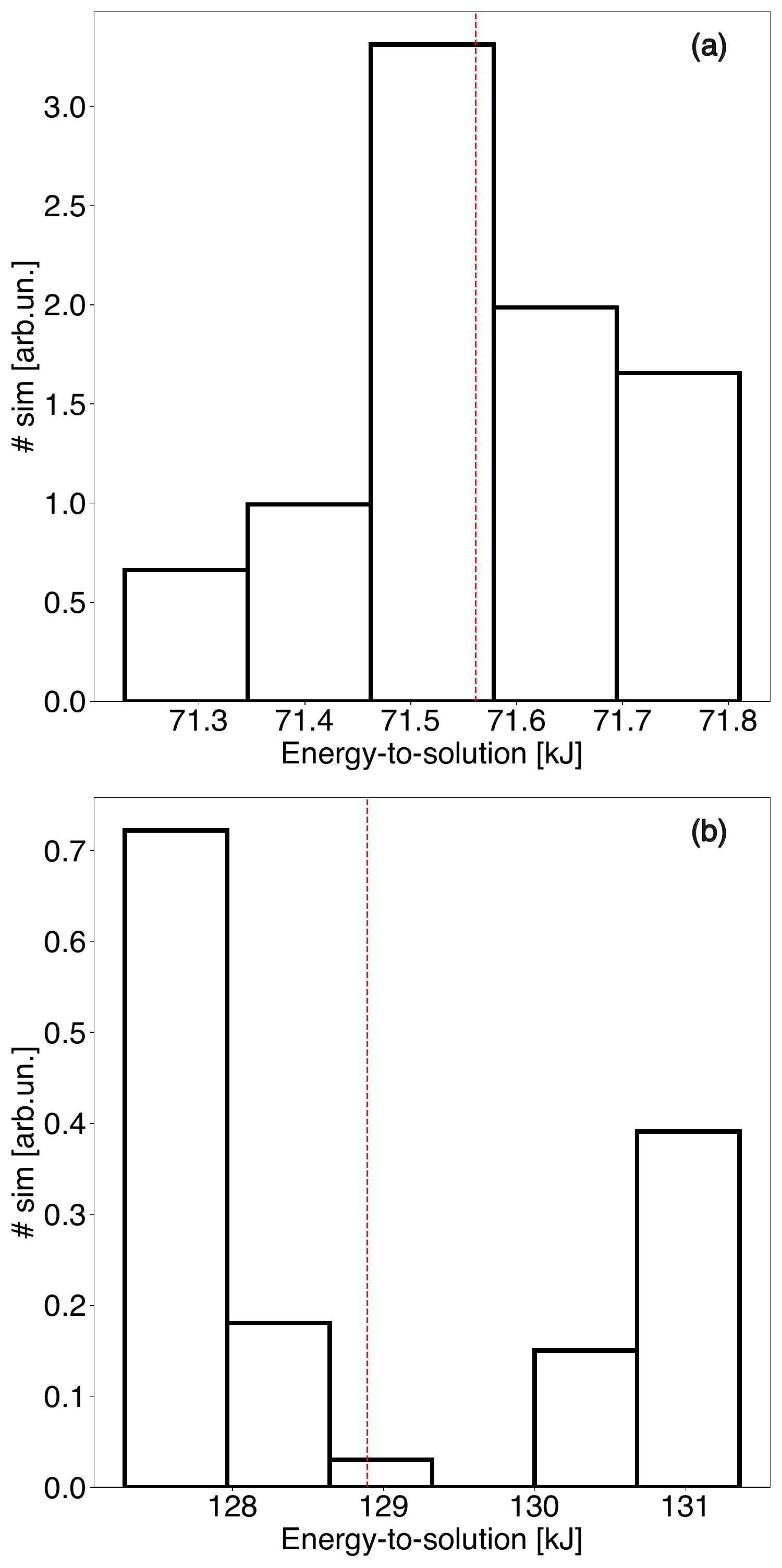}
    \vspace{-4mm}
    \caption{Distribution of energy values across simulations for runs on device + CPU (a) and on CPU only (b). The read dashed lines represent the average energy-to-solution.}
    \label{fig:energy_distribution}
\end{figure}

Starting from power measurements, the energy-to-solution for each Wormhole card is calculated as the discrete integral of power over the simulation time (excluding the \verb|sleep| phases). These values are then summed to obtain the total energy consumption of all cards.
The energy consumed by the dual-socket processor is computed by summing the values recorded with \verb|perf|, again considering only the actual simulation time.
We compute the energy-to-solution required for a simulation as the sum of the energy consumed by the Wormhole cards and the processor. 
Fig.~\ref{fig:energy_distribution} shows the energy distribution for the same simulations of Fig.~\ref{fig:time_distribution}.
On average, the accelerated simulations consume $71.56 \pm 0.13 \, \mathrm{kJ}$ with values ranging from $71.23$ to $71.81 \, \mathrm{kJ}$.
In contrast, the classical simulations require significantly more energy, with an average energy-to-solution of $128.89 \pm 1.52 \, \mathrm{kJ}$, and values ranging from $127.29$ to $131.36 \, \mathrm{kJ}$.
The slightly higher standard deviation observed in the classical case is likely attributable to the variability in runtime behavior. Executing the code on the Tenstorrent platform results in a $1.80 \times$ reduction in energy-to-solution. It is worth noting that this energy saving comes with a modest increase in peak power consumption. Specifically, the maximum power drawn by the CPUs and cards during execution of the accelerated code reaches $\approx 260 \mathrm{W}$, compared to $\approx 210\, \mathrm{W}$ for the reference code. 

\section{Summary and perspectives}\label{sec:conclusion}
In this work, we demonstrated the first (to the best of our knowledge) porting of an astrophysical $N$-body code to the RISC-V-based Tenstorrent Wormhole n300 accelerator. Our performance evaluation, in terms of time-to-solution and energy-to-solution, shows that the accelerated code delivers over $2 \times$ faster execution and about $2 \times$ higher energy efficiency compared to a highly optimized, OpenMP-parallelized CPU version exploiting AVX-512 instructions. These results highlight the potential of RISC-V-based accelerators as a competitive, energy-efficient solution for next-generation astrophysical simulations employing this class of algorithms.

As an immediate next step, we aim to implement the force kernel in CUDA in order to compare the performance of the Tenstorrent Wormhole n300 accelerator with that of a conventional NVIDIA GPU. This will also allow us to study the effect of increasing the number of particles to assess suitability in real HPC contexts.
Subsequently, we plan to extend our benchmarks to MPI with multiple accelerators and to modify and optimize the code to fully exploit card-level parallelism, which ultimately will enable us to perform both strong and weak scalability tests.


\begin{acks}
We thank E. Duffy, R. Friedman, R. Ganisetti, and F. LeClair (Tenstorrent) for their guidance; F. Proverbio and A. D’Apice (E4) for technical support; and F. Magugliani (E4) for comments on the manuscript.

This research is supported
 by the Italian Research Center on High Performance Computing Big Data and Quantum Computing (ICSC), project funded by European Union - NextGenerationEU - and National Recovery and Resilience Plan (NRRP) - Mission 4 Component 2 within the activities of Spoke 3
 (Astrophysics and Cosmos Observations).
\end{acks}

\bibliographystyle{ACM-Reference-Format}
\bibliography{risc-v4astro}

\end{document}